\documentclass[twocolumn]{aastex63}
\newcommand{\todo}{\ifmmode \text{\color{purple}\Huge{\(\bullet\)}} \else {\color{purple}{\Huge$\bullet$}}\fi}
\newcommand{\finish}{\ifmmode \text{\color{blue}\Huge{\(\bullet\)}} \else {\color{blue}{\Huge$\bullet$}}\fi}

\newcommand{\Msun}{M_{\odot}}
\newcommand{\Mbh}{M_\mathrm{BH}}

\newcommand{\msun}{M_{\odot}}

\newcommand{\lambdaedd}{\lambda_\mathrm{Edd}}

\newcommand{\R}{\mathcal{R_{\rm rest}}}

\newcommand{\whz}{\mathrm{W}~\mathrm{Hz}^{-1}}

\usepackage{natbib}
\usepackage{amsmath}
\usepackage{multirow} 
\usepackage{ifthen} 

\received{May 15, 2026}
\accepted{June 15, 2026}
\submitjournal{ApJL}

\shorttitle{VLBA jet in a super-Eddington quasar}
\shortauthors{Obuchi et al.}

\begin{document}



\title{
A VLBA-resolved Jet Associated with Super-Eddington Accretion in a Radio-loud Quasar at $z=3.4$}

\correspondingauthor{Sakiko Obuchi}
\email{buchi-13526.cjl@fuji.waseda.jp}

\author[0009-0007-6567-4240]{Sakiko Obuchi}
\affiliation{
Department of Physics, School of Advanced Science and Engineering, Faculty of Science and Engineering, Waseda University, 3-4-1,
Okubo, Shinjuku, Tokyo 169-8555, Japan}

\author[0000-0002-5208-1426]{Ingyin Zaw}
\affiliation{New York University Abu Dhabi, Science Division, Physics Program, P.O. Box 129188, Abu Dhabi, UAE}

\author[0000-0001-6906-772X]{Kazuhiro Hada}
\affiliation{Graduate School of Science, Nagoya City University, Yamanohata 1, Mizuho-cho, Mizuho-ku, Nagoya 467-8501, Aichi, Japan}
\affiliation{Mizusawa VLBI Observatory, National Astronomical Observatory of Japan, 2-12 Hoshigaoka, Mizusawa, Oshu, Iwate 023-0861, Japan}

\author[0000-0002-4377-903X]{Kohei Ichikawa}
\affiliation{Frontier Research Institute for Interdisciplinary Sciences, Tohoku University, Sendai, Miyagi 980-8578, Japan}
\affiliation{Astronomical Institute, Tohoku University, Aramaki, Aoba-ku, Sendai, Miyagi 980-8578, Japan}

\author[0000-0003-4679-1058]{Joseph D. Gelfand}
\affiliation{New York University Abu Dhabi, Science Division, Physics Program, P.O. Box 129188, Abu Dhabi, UAE}



\begin{abstract}
We report the detailed jet properties of eROSITA Final Equatorial Depth Survey (eFEDS) J084222.9+001000 (hereafter ID830), a radio-loud super-Eddington quasar at $z=3.4351$, revealed by Very Long Baseline Array (VLBA) observations at 1.6~GHz, 4.9~GHz, and 8.2~GHz. Thanks to the high spatial resolution of the VLBA, we successfully resolve a parsec-scale core--jet structure of ID830, and find a well-collimated jet extending over $\approx 745$~pc, making it the most distant and one of the very few currently known radio-loud quasars with a resolved jet associated with super-Eddington accretion. The physical scale and evolutionary track of ID830 differs markedly from the low-$z$ analogues, such as nearby radio-luminous high-Eddington narrow-line Seyfert 1 galaxies, suggesting that this source represents a distinct high-$z$ population compared to previously known samples, with important implications for AGN feedback in early galaxy evolution.
We also find that the jet has a relativistic speed of $v \gtrsim 0.19c$ and a modest viewing angle of $\phi \lesssim 79^\circ$ to the line of sight, although its emission is not significantly Doppler-boosted ($\delta \sim 1$). This provides the first evidence that such a relativistic and collimated jet can be produced over several hundred parsecs in the super-Eddington phase, lasting for at least $10^{3\text{--}4}$~yr. Our results call for further theoretical and numerical studies to understand the physical processes required to sustain such large-scale collimation in super-Eddington accretion, which remains a missing piece.
\end{abstract}




\section{Introduction}\label{sec:intro}
Relativistic jets launched from supermassive black holes (SMBHs) are a key activity for understanding the growth of SMBHs and their co-evolution with host galaxies. Such jet-hosting sources are called radio active galactic nuclei (AGN), and they account for roughly 10\% of the overall AGN population \citep{Kellermann1989,Ivezic2002}. Their powerful jets can suppress (or enhance) star formation through heating and ejection of the surrounding gas, commonly referred to as ``AGN feedback", which plays a crucial role in cosmic evolution of galaxy and SMBH growth \citep[e.g.,][]{Best2005,McNamara2007,Fabian2012}. 

Although the physical mechanism of jet formation remains an open question in modern astronomy, several studies have revealed a strong connection with the accretion state. At high accretion rates, geometrically thin and optically thick standard disks \citep{Shakura1973} are radiatively efficient and tend to suppress jet production (radiative-mode AGN), whereas at low accretion rates, geometrically thick and optically thin radiatively inefficient accretion flows \citep[RIAF;][]{Ichimaru1977,Narayan1994} are capable of launching powerful, highly collimated jets \citep[jet-mode AGN, e.g.,][]{Ho2008,Heckman2014}. 
Observations further revealed an inverse correlation between radio loudness (radio-to-optical luminosity ratio) and the Eddington ratio, indicating that jet production favors lower accretion rates \citep[e.g.,][]{Ho2002,Sikora2007}.

Accretion onto SMBHs is not limited to sub-Eddington regimes. 
Theoretically, super-Eddington accretion has been shown to occur under certain conditions \citep[e.g.,][]{Abramowicz1988,Ohsuga2009,Inayoshi2016,Takeo2020}. Observational candidates undergoing such super-Eddington accretion have also been reported in various systems, including ultra-luminous X-ray sources \citep[ULXs;][]{Kaaret2017}, tidal disruption events \citep[TDEs;][]{Bloom2011,Burrows2011}, narrow-line Seyfert 1 galaxies \citep[NLS1s;][]{Collin2004,Yang2020}, and even high-$z$ quasars \citep{Harikane2023,Maiolino2024,Suh2025}. Several theoretical models have suggested jet formation in the super-Eddington regime driven by either radiation pressure or magnetic fields \citep[e.g.,][]{Ohsuga2011,Sadowski2015,Jiang2019}, but observational tests remain lacking since both super-Eddington accretion and radio AGNs are intrinsically limited in number.

Recently, \cite{Ichikawa2023} constructed a catalog of radio AGN\footnote{In the catalog, a radio-loudness cut of $\log R_\mathrm{obs} > 10$ \citep{Kellermann1989,Ivezic2002} was applied to the sample in order to focus on radio emission originating from AGN--jet activity rather than star formation. The observed radio loudness is defined as $\log R_\mathrm{obs} = \log(F_\mathrm{int}/F_{i\mathrm{\text{-}band}})$, where $F_\mathrm{int}$ is the total integrated flux density in the VLA/FIRST band and $F_{i\mathrm{\text{-}band}}$ is the cmodel flux density in the Subaru/Hyper Suprime-Cam (HSC) band \citep{Ichikawa2023}.} in the eROSITA Final Equatorial Depth Survey (eFEDS) field \citep{Predehl2021,Brunner2022,Merloni2024} by cross-matching the VLA/FIRST \citep{Becker1994,Helfand2015} and eROSITA X-ray source catalogs \citep{Liu2022,Salvato2022}. Among them, \cite{Obuchi2026} reported one particularly interesting radio-loud ($\log R_\mathrm{obs}=3.02$), super-Eddington quasar at $z=3.4351$, eFEDS~J084222.9+0010000 (cataloged as eFEDS object ID=830; hereafter ID830). ID830 has a black hole mass of $\Mbh = (4.40 \pm 0.72) \times 10^{8} \msun$ and exhibits super-Eddington accretion, with $\lambda_\mathrm{Edd,UV}=1.44 \pm 0.24$ from the UV luminosity and $\lambda_{\mathrm{Edd,X}} = 12.8 \pm 3.9$ from the X-ray luminosity \citep{Obuchi2026}, making it a rare super-Eddington quasar reaching 1.4~GHz luminosity of $L_\mathrm{1.4GHz}=10^{27.1}$~$\whz$, suggestive of a radio jet \citep[e.g.,][]{Mauch2007}.

In this letter, we present Very Long Baseline Array (VLBA) observations of ID830. With the milliarcsecond (mas) high resolution of the VLBA, we find that ID830 hosts a collimated radio jet extending from pc to nearly kpc scale. We discuss the implications for jet formation in the super-Eddington accretion, which has remained largely unexplored.
We adopt the following cosmological parameters throughout this paper: 
$H_0 = 68$\,km\,s$^{-1}$\,Mpc$^{-1}$, $\Omega_\mathrm{M}=0.31$, and $\Omega_\Lambda=0.69$ \citep{Planck2020}.

\section{Observations and Data Reduction}\label{sec:data_reduction}

The VLBA observations of ID830 were conducted on 2025 March 31, under the project code BZ121 (PI: I. Zaw). We observed the target at three frequencies, 1.6~GHz (L-band), 4.9~GHz (C-band), and 8.2~GHz (X-band). The total observing time was 5.6~hr with 3.6~hr on the target (1.2~hr per band). The observations for the target were performed in a single session alternating between the L, C, and X bands at 30 minute intervals to maximize the $uv$ coverage and the dynamic range of the image. 
For each band, the signals were recorded using the digital downconverter (DDC) algorithm with 2-bit quantization, and the total bandwidth of 512~MHz was divided into four intermediate frequency channels (IFs) of 128~MHz each, in dual (right- and left-hand) circular polarization. The data were also correlated with the VLBA DiFX correlator (version 2.8.1) in Socorro, New Mexico \citep{Deller2011}, with a correlator integration time of 2~s. Since the target was expected to be rather weak, we utilized the phase-referencing technique \citep{Beasley1995}, where the observations involved regular switching between the target and the nearby bright phase calibrator J0839+0104 separated by 1\fdg11 on the sky, plus a few additional scans of the fringe finder DA193. Although all ten antennas were scheduled for the observations, the Hancock (HN) antenna was not used due to an electronic failure. 


\begin{figure*}[tp!]
\begin{center}
\includegraphics[width=0.99\textwidth]{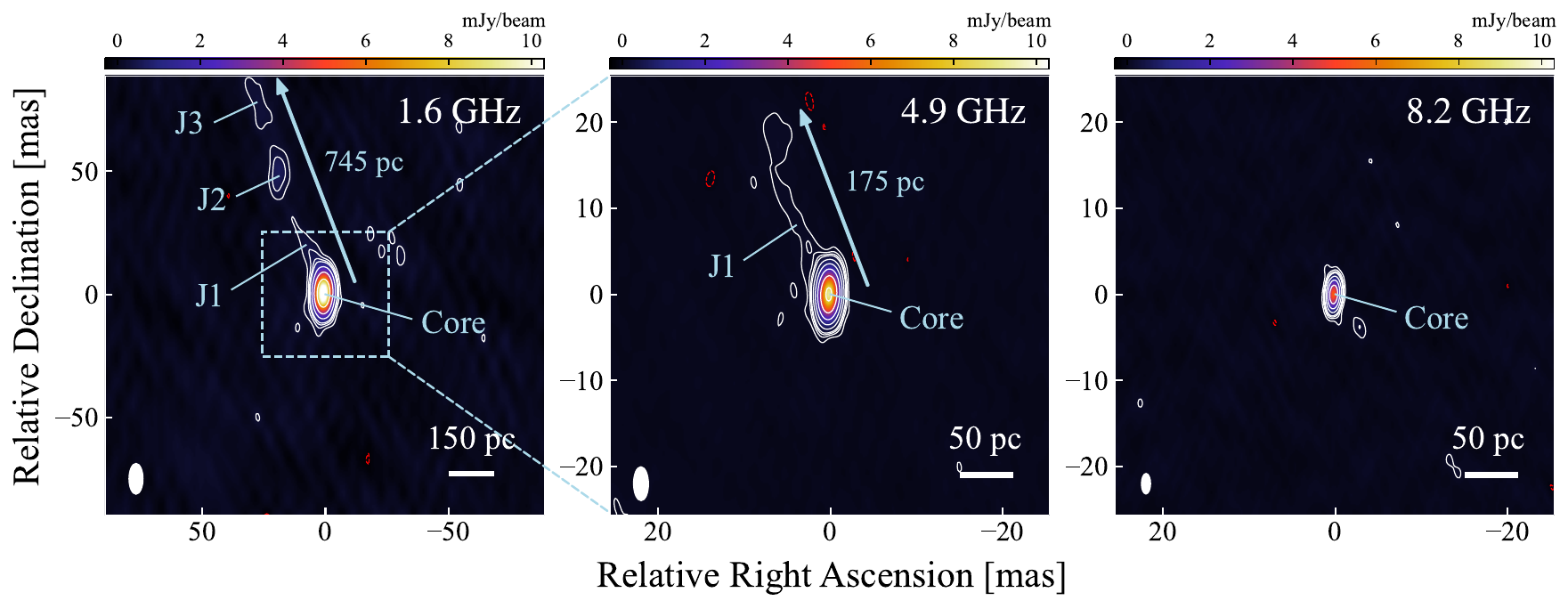}
\caption{
The VLBA images of ID830 at 1.6~GHz (left), 4.9~GHz (middle), and 8.2~GHz (right) with natural weighting. The negative and positive contours are shown as the red dashed and solid white lines, respectively. The contour levels are at $-1$, 1, 2, 3, 4, 8, ..., $2^n$ times the $3 \sigma$ rms noise level. The white ellipse in each panel represents the beam size. The core and jet components (J1--J3) are also labeled in the figure.
}\label{fig:VLBAimages}
\end{center}
\end{figure*}

Our data reduction was performed using the National Radio Astronomy Observatory Astronomical Image Processing System \citep[AIPS;][]{Greisen2003} following standard VLBI data reduction procedures. We firstly calibrated the data of J0839+0104 and the derived phase solutions were transferred to the target. The L-band data were partially flagged in AIPS because of radio frequency interference (RFI). Imaging and self-calibration were performed in the DIFMAP software \citep{Shepherd1994}. 
Since the phase calibrator J0839+0104 showed a core with knotty jet structure, the observed visibility phases of the calibrator included phase contributions from its intrinsic source structure in addition to atmospheric and instrumental phase errors. We therefore additionally subtracted the visibility phases associated with the 
calibrator structure from the calibration solutions before applying them to the target, which significantly improved the quality of the phase-referenced maps of the target \citep[see also][]{Beasley1995}. 
As the present study focuses on detailed imaging of the target rather than astrometric aspects, self-calibration of the target source itself was also carried out in phase and amplitude at L- and C-band, and in phase only at X-band, given the sufficiently high signal-to-noise ratio (S/N). The solution intervals were gradually shortened, reaching 120 min for the L-band amplitude self-calibration, 60 min for the C-band amplitude self-calibration, and 30 min for the X-band phase calibration. The amplitude gain corrections were stable within $\approx$ 20\% across the array, ensuring a conservative and reliable calibration.

\section{Results}\label{sec:results}

\renewcommand{\arraystretch}{1}
\begin{deluxetable}{cccccc}
  \tablecaption{Summary of VLBA Observations.}\label{tab:VLBAobsparams}
  \tablehead{
       \colhead{Freq.} & \colhead{$B_{\rm{maj}}$} & \colhead{$B_{\rm{min}}$} & 
       \colhead{$B_{\rm{p.a.}}$} & \colhead{$S_{\rm{peak}}$} & \colhead{rms}    
       \\ [-4pt]
        \colhead{(GHz)} & \colhead{(mas)} & \colhead{(mas)} & \colhead{(deg)} & \colhead{(mJy/bm)} & \colhead{(mJy/bm)}
        \\ [-4pt]
         \colhead{\footnotesize(1)} & \colhead{\footnotesize(2)} & \colhead{\footnotesize(3)} & \colhead{\footnotesize(4)} & \colhead{\footnotesize(5)} & \colhead{\footnotesize(6)}
     }
    \startdata
1.63 & 12.2 & 5.41 & 
$-0.01$ &
12.6 & 0.073 \\
4.88 & 3.85 & 1.65 & 
0.47 &
9.26 & 0.022 \\
8.17 & 2.29 & 0.98 & 
$-0.40$ &
5.69 & 0.035 \\
    \enddata
\tablenotetext{}{
Notes. Columns: (1) Central frequencies. (2) Major axis FWHM of the synthesized beam. (3) Minor axis FWHM of the synthesized beam. (4) Position angle of the major axis of the synthesized beam. (5) Peak flux density. (6) $1 \sigma$ rms image noise level.}
\end{deluxetable}

\renewcommand{\arraystretch}{1}
\setlength{\tabcolsep}{14pt} 
\begin{deluxetable*}{ccccccc}
\thispagestyle{empty}
  \tablecaption{Summary of Model-fit Parameters and Results.}\label{tab:VLBAfitparams}
  \tablehead{
        \colhead{Freq.} & \colhead{$\theta_{\rm{maj}}$} & \colhead{$\theta_{\rm{min}}$} & 
       \colhead{$\theta_{\rm{p.a.}}$} & \colhead{$S_{\rm{int,core}}$} & \colhead{$S_{\rm{int,jet}}$} & \colhead{$T_{\rm{b}}$}
       \\ [-4pt]
        \colhead{(GHz)} & \colhead{(mas)} & \colhead{(mas)} & \colhead{(deg)} & \colhead{(mJy)} & \colhead{(mJy)} & \colhead{($10^{10}$\,K)}
        \\ [-4pt]
        \colhead{\footnotesize(1)} & \colhead{\footnotesize(2)} & \colhead{\footnotesize(3)} & \colhead{\footnotesize(4)} & \colhead{\footnotesize(5)} & \colhead{\footnotesize(6)} & \colhead{\footnotesize(7)}
     }
    \startdata
1.63 & $3.27 \pm 0.25$ & $<0.58$ & 
$-23.0$ &
$13.42 \pm 1.94$ & $0.82 \pm 0.08$ & $>1.44$\\
4.88 & $1.26 \pm 0.06$ & $<0.12$ & 
$-27.1$ &
$10.23 \pm 1.22$ & $0.37 \pm 0.04$ & $> 1.60$ \\
8.17 & $1.12 \pm 0.09$ & $0.30 \pm 0.02$ & 
$-14.7$ &
$6.86 \pm 0.98$ & $< 0.40$ & $0.17 \pm 0.03$ \\
    \enddata
\tablenotetext{}{
Notes. Columns: (1) Central frequencies. (2) Major axis FWHM of the elliptical Gaussian model. (3) Minor axis FWHM of the elliptical Gaussian model. (4) Position angle of the major axis of the elliptical Gaussian model. (5) Integrated flux density of the core. (6) Integrated flux density of the jet (J1). (7) Core brightness temperature at the rest-frame frequency.}
\end{deluxetable*}

\subsection{VLBA Images}\label{sec_sub:images}

Figure~\ref{fig:VLBAimages} shows the VLBA images of ID830 at the three observed frequencies, 1.6~GHz, 4.9~GHz, and 8.2~GHz, all created with natural weighting in DIFMAP. Each parameter of ID830 obtained from the observations is summarized in Table~\ref{tab:VLBAobsparams}. In Figure~\ref{fig:VLBAimages}, the core of the target is clearly detected at all frequencies. In addition, a prominent one-sided core--jet structure is detected at both 1.6 and 4.9~GHz, elongating toward the northeast with a position angle of $\mathrm{PA}\approx 20^\circ$ (from North to East) in both bands. At 1.6~GHz, the jet is observed as three separate components (J1, J2, and J3 in Figure~\ref{fig:VLBAimages}), which are aligned along the same direction up to $\approx 745$~pc from the core. At 4.9~GHz, part of the jet seen at 1.6~GHz (component J1) appears as a well-collimated structure, extending over $\approx 175$~pc. Although the jet structure is weak, the target self-calibration described in Section~\ref{sec:data_reduction} significantly improved the rms noise level and dynamic range by a factor of 2--6, depending on the observing band. The jet component was already detected at both 1.6 and 4.9~GHz after phase-only self-calibration. Most importantly, the consistent jet structure is independently detected at these two frequencies, strongly indicating that the extended emission is not an artifact. We also create $uv$-tapered images at 1.6 and 4.9~GHz using the \verb|uvtaper| task in DIFMAP to further confirm the robustness of the jet detection by increasing sensitivity to low-surface-brightness emission (see \hyperref[app:tapered_images]{Appendix} for details).
At 8.2~GHz, no distinct jet structure is detected, which is consistent with the expectation that emission from a diffuse, low-density region such as a lobe-like structure, has a steeper spectral index than that of the core, as discussed in Section~\ref{sec_sub:radioSED}. We note that the $uv$ coverage may also affect the sensitivity to extended structures, since higher-frequency observations can only detect emission from more compact structures.

\begin{figure}[tp!]
\begin{center}
\includegraphics[width=0.48\textwidth]{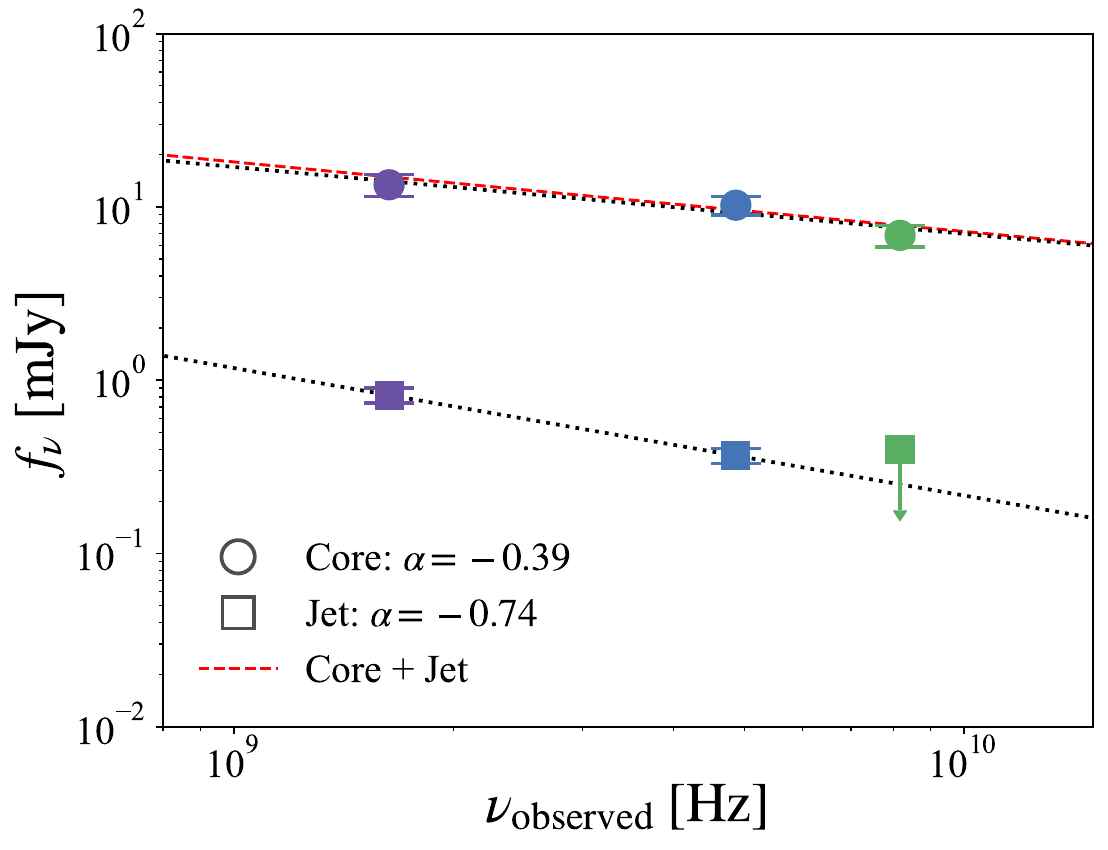}
\caption{
The observed-frame radio SED of ID830. The circles and squares indicate the flux densities of the core and jet (J1) components at 1.6~GHz (purple), 4.9~GHz (blue), and 8.2~GHz (green), respectively. The black dotted lines represent the best-fit spectral slopes of the core and jet components, respectively. The red dashed line represents the best-fit slope of the total (core+jet) component.
}\label{fig:radioSED}
\end{center}
\end{figure}

\begin{figure}[tp!]
\begin{center}
\includegraphics[width=0.48\textwidth]{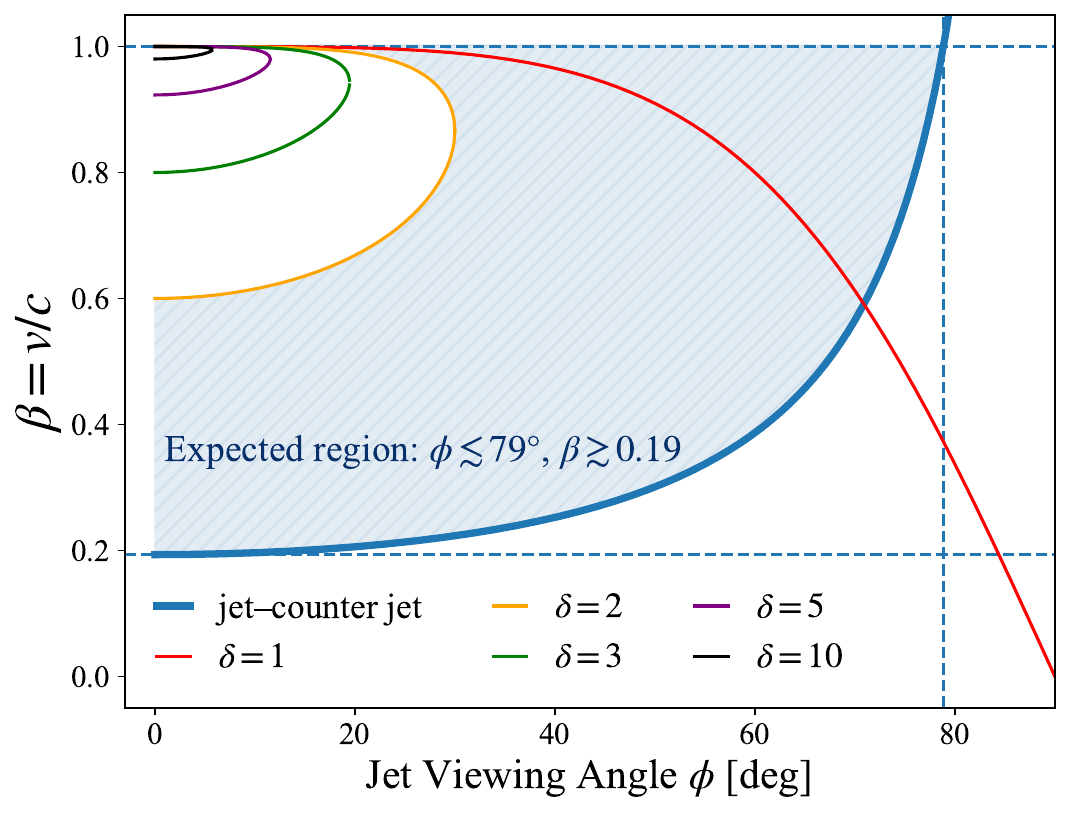}
\caption{
Jet viewing angle $\phi$ vs. intrinsic jet speed $\beta$. The blue solid curve shows the lower limit of the $\beta$--$\phi$ relation derived from the constraint $J \gtrsim 2.92$, and the other solid curves (red, orange, green, purple and black) correspond to Doppler factors ranging from 1 to 10. The shaded region represents the allowed parameter space satisfying the constraints on $\phi$ and $\beta$, i.e., $J\gtrsim 2.92$ and $\delta \sim 1$ ($\delta < 2$).}
\label{fig:jet_speed}
\end{center}
\end{figure}

We also estimate the integrated flux density of the central core of ID830 at each frequency by fitting an elliptical Gaussian model using the \verb|modelfit| task in DIFMAP.
The resulting core flux densities are $13.42 \pm 1.94$~mJy at 1.6~GHz, $10.23 \pm 1.22$~mJy at 4.9~GHz, and $6.86 \pm 0.98$~mJy at 8.2~GHz, respectively. For the jet components, we estimate the flux densities directly from the \verb|CLEAN| images using the same region across the three bands after matching the pixel size and resolution to those of the 1.6~GHz image, to ensure a consistent comparison. The region is defined based on the jet morphology of the J1 component at 4.9~GHz, which has the highest S/N. At 8.2~GHz, the jet flux density is reported as a $3 \sigma$ upper limit, estimated over the jet region defined from the lower-frequency image, because no extended jet emission is detected above the $3 \sigma$ level. 
The uncertainties of the model-fit parameters are estimated following the widely used method of \cite{Fomalont1999} and \cite{Lee2008}. Since the minor axis FWHMs of the Gaussian fitting at 1.6~GHz and 4.9~GHz are comparable to the minimum resolvable size estimated by the formula provided by \cite{Lee2008} and \cite{Lobanov2005}, the component is considered unresolved and the minimum resolvable size is used as an upper limit in the subsequence analysis. 
Additionally, we adopt a typical 10\% uncertainty, dominated by the VLBA amplitude calibration based on the antenna system temperatures and gain curves \citep[e.g.,][]{An2012,Cao2017,Hada2018}. For the core flux densities, the uncertainties are estimated by adding 10\% errors in quadrature to the formal fitting errors. The model-fit parameters and best-fit results are summarized in Table~\ref{tab:VLBAfitparams}.

\subsection{Radio Spectral Properties}\label{sec_sub:radioSED}

Based on the estimated flux densities, we plot the radio spectral energy distribution (SED) of ID830 for both the core and jet components in Figure~\ref{fig:radioSED}. 
The spectra of ID830 are well described by a power-law function $S_\nu \propto \nu^{\alpha}$, consistent with the nonthermal radio continuum emission. The core component typically shows an optically thick or self-absorbed flat spectrum with $\alpha \gtrsim -0.5$, and the jet component shows an optically thin steep spectrum with $\alpha \lesssim -0.5$ \citep[e.g.,][]{Padovani2017,Blandford2019}. 
In this study, the spectral indices of the core and jet (J1) are measured to be $\alpha_\mathrm{core}=-0.39 \pm 0.14$ (between all three bands) and $\alpha_\mathrm{jet} = -0.74 \pm 0.13$ (between 1.6 and 4.9~GHz), respectively, indicating that ID830 exhibits a flat-spectrum core and a steep-spectrum jet. Since the upper limit on the jet flux density at 8.2~GHz does not provide a strong constraint because of the relatively high rms noise level, the jet spectral index is derived using only two lower-frequency measurements. 
Figure~\ref{fig:radioSED} also presents the spectrum derived from the total flux densities, including both the core and jet (J1) components (red dashed line). The moderately flat total spectral index of $\alpha_\mathrm{total} \approx -0.40$ and the much brighter core emission relative to the jet emission suggest that the radio emission of ID830 is core-dominant.

The total (core+jet) radio spectrum shown in Figure~\ref{fig:radioSED} is consistent with the FIRST 1.4~GHz flux density \cite[16.7~mJy;][]{Obuchi2026}, obtained with an angular resolution of $5 \arcsec$ \citep{Becker1994,Helfand2015}, within a factor of $\sim 1.1$. 
This suggests that any radio component on scales larger than the $\approx 745$~pc VLBA-resolved structure does not contribute substantially to the total 1.4~GHz flux.
In other words, the current data do not require a significantly more extended, flux-dominant jet component beyond the VLBA-resolved structure shown in Figure~\ref{fig:VLBAimages}. 
In contrast, the Very Large Array Sky Survey \cite[VLASS;][]{Lacy2020} 3~GHz flux density \citep[24.4~mJy;][]{Obuchi2026}, obtained with an angular resolution of $2.5 \arcsec$, is brighter than the VLBA radio SED by a factor of $\sim 2.1$. Given the consistency between the FIRST and VLBA flux densities, this discrepancy likely reflects intrinsic variability of the core flux, which can strongly affect the observed radio flux in such core-dominant jets.

\subsection{Relativistic Beaming and Jet Properties}\label{sec_sub:jet_properties}

The brightness temperature is a key diagnosis of radio emission, reflecting the source intensity and providing constraints on the relativistic beaming factor. We calculate the core brightness temperature $T_\mathrm{b}$ in the rest-frame following the equation given by \cite{Kovalev2005}:
\begin{equation}
    T_\mathrm{b} = 1.22 \times 10^{12}\,\frac{S_\mathrm{int,core} (1+z)}{\theta_\mathrm{maj} \theta_\mathrm{min} \nu^2} ~\mathrm{[K]},
\end{equation}
where $S_\mathrm{int,core}$ is the integrated core flux density in Jy, $\theta_\mathrm{maj}$ and $\theta_\mathrm{min}$ are the fitted FWHM of the elliptical Gaussian model in mas, and $\nu$ is the observing frequency in GHz. The estimated brightness temperatures of ID830 at all frequencies listed in Table~\ref{tab:VLBAfitparams} exceed $T_\mathrm{b}=10^{10}$~K, which is well above the upper limit expected for radio emission from normal star formation ($T_\mathrm{b} \lesssim 10^5$~K), and are consistent with nonthermal emission from AGN jets  \citep[$T_\mathrm{b} \gtrsim 10^6$~K; ][]{Kellermann1969,Condon1992,Kovalev2005}.
Such high brightness temperatures have a physical upper limit at $T_\mathrm{b} \sim 10^{12}$~K due to the inverse Compton catastrophe, and a more realistic limit is given by the equipartition brightness temperature, $T_\mathrm{b,eq} \approx 5 \times 10^{10}$~K \citep{Readhead1994}. When the observed brightness temperature exceeds the intrinsic value, the radio emission can be interpreted as being strongly enhanced by relativistic beaming, with the Doppler factor $\delta = T_\mathrm{b} / T_\mathrm{b,int}$. Although the intrinsic brightness temperature $T_\mathrm{b,int}$ is often assumed as $T_\mathrm{b,eq}$, we conservatively adopt a somewhat smaller value of $T_\mathrm{b,int} \approx 4.1 \times 10^{10}$~K \citep{Homan2021}, measured from the large sample of parsec-scale core bright jets. 
The observed core brightness temperatures listed in Table~\ref{tab:VLBAfitparams} are well below $T_\mathrm{b,int}$ at the highest frequency, and the lower limits at the other two frequencies are also comparable to $T_\mathrm{b,int}$, suggesting little or no significant Doppler boosting ($\delta \sim 1$) in the radio emission of ID830\footnote{Note that the intrinsic brightness temperature may be lower at high rest-frame frequencies, which could lead to an underestimation of the Doppler factor \citep{Lee2014}; however, our conclusions remain unchanged even in such a case.}. 
This is consistent with the non-blazar nature of ID830 \citep{Obuchi2026}, as evidenced by a steep X-ray photon index ($\Gamma = 2.43$) and moderate dust extinction ($A_V = 0.39$~mag).

\begin{figure*}[tp!]
\begin{center}
\includegraphics[width=0.9\textwidth]{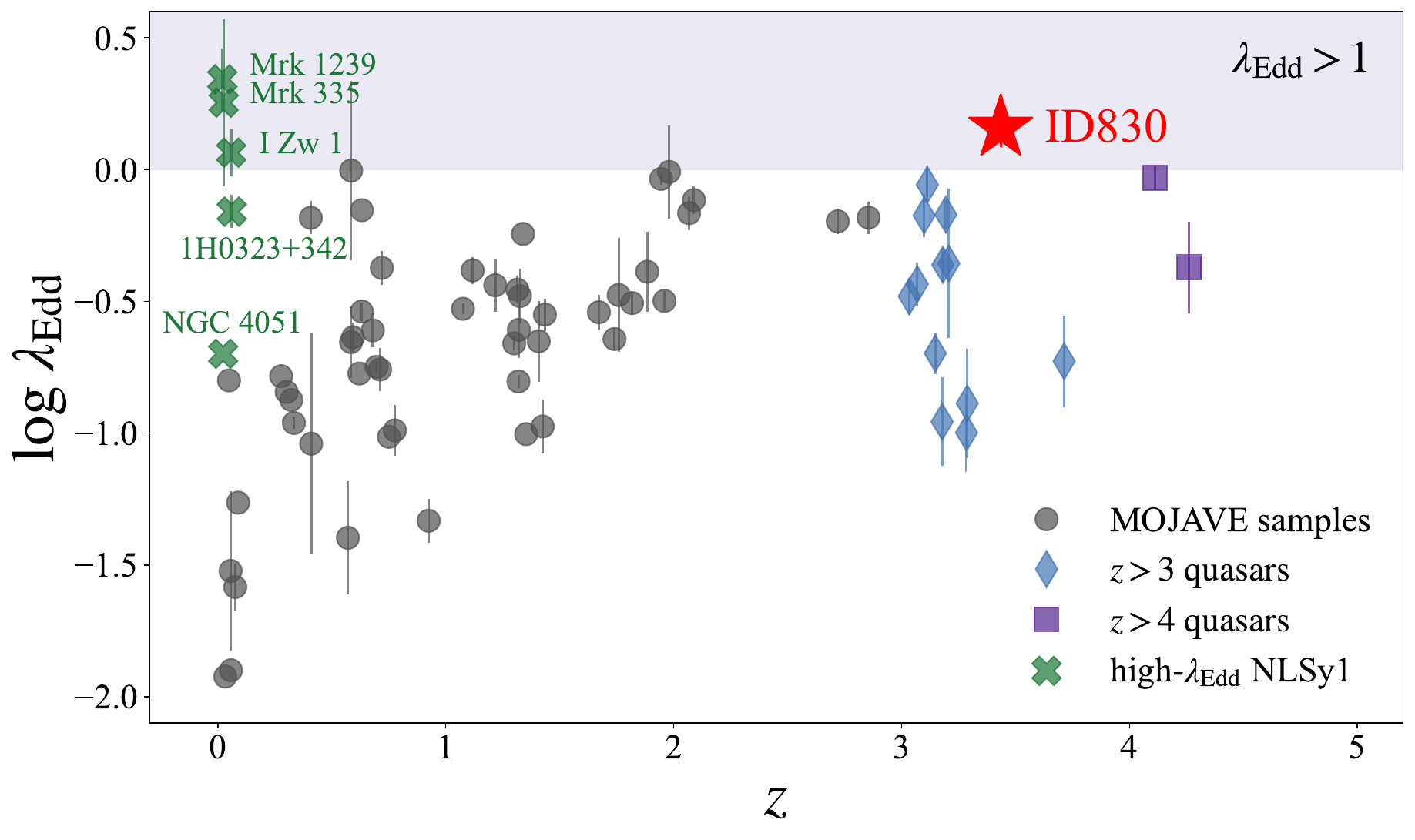}
\caption{
Redshift $z$ vs. Eddington ratio $\lambdaedd$. All sources shown in the figure are non-blazar objects with parsec-scale resolved core--jet structures and well-determined Eddington ratios. 
ID830 is shown as a red star, lying within the shaded region corresponding to the super-Eddington regime ($\lambdaedd > 1$).
The gray circles, blue diamonds, and purple squares indicate the MOJAVE sample \citep{Lister2018}, $z>3$ quasars \citep{Sotnikova2021,Sotnikova2024}, and $z>4$ quasars \citep{Krezinger2022}, respectively.
The Eddington ratios are compiled from the following studies: SDSS DR14 \citep{Rakshit2020}, SDSS DR16 \citep{Wu2022},  \cite{Chai2012}, \cite{Sikora2007}, and \cite{Tadhunter2003}. For $z>2$ samples with \ion{C}{4}-based $\Mbh$ and $\lambdaedd$ estimates, we apply individual corrections based on the \ion{C}{4} blueshift velocities \citep{Coatman2017}, since \ion{C}{4}-based measurements are often biased due to strong outflows. Additionally, some high-$\lambdaedd$ NLS1s are plotted as green cross markers (Mrk 1239: \citealt{Doi2015,Pan2021}; Mrk 335: \citealt{Yang2020,Yao2021,Wang2023a,Wang2023b}; I Zw 1: \citealt{Wilkins2017,Wang2023a,Wang2023b,Yang2024}; 1H 0323+342: \citealt{Landt2017,Kynoch2018,Hada2018}; NGC 4051: \citealt{Denney2009,Giroletti2009,Yuan2021}).
}\label{fig:z_vs_lambdaEdd}
\end{center}
\end{figure*}

We can further constrain the jet speed and viewing angle using the observed jet morphology and the inferred Doppler factor. 
Relativistic beaming affects the jet morphology, particularly the appearance of the counter jet, because Doppler boosting enhances the flux of the approaching jet, making the counter jet appear fainter in comparison. Following \cite{Ghisellini1993}, the ratio of the jet flux to the counter jet flux, $J$, can be expressed as:
\begin{equation}\label{eq_jetratio}
    J = \biggl( \frac{1 + \beta \cos{\phi}}{1 - \beta \cos{\phi}} \biggr)^{2-\alpha},
\end{equation}
where $\beta=v/c$ is the intrinsic jet speed, $\phi$ is the jet viewing angle to the line-of-sight, and $\alpha$ is the jet spectral index. Based on the 4.9~GHz image, we estimate a lower limit on the jet--counter jet flux ratio of $J \gtrsim 2.92$. Since the counter jet is undetected in ID830, we measure its counter jet flux in a region symmetric to the detected jet across the core, adopting a $1 \sigma$ upper limit. 
By substituting the lower limit $J \gtrsim 2.92$ and the jet spectral index $\alpha=-0.74$ into Equation~(\ref{eq_jetratio}), we obtain the expected region in the $\beta$--$\phi$ plane, yielding constraints of $\phi \lesssim 79^\circ$ and $\beta \gtrsim 0.19$, as shown in Figure~\ref{fig:jet_speed}. Figure~\ref{fig:jet_speed} also reflects the constraints on the Doppler factor, excluding the region with $\delta \ge 2$ where Doppler boosting may be significant, since ID830 has a Doppler factor of $\delta \sim 1$ ($\delta < 2$). Our results imply that the outflow is not a simple wind-like component but a collimated relativistic jet, with a bulk speed of at least $0.1c$--$0.2c$. At the same time, we argue that ID830 is unlikely to be a strongly Doppler-boosted blazar, but rather a misaligned AGN viewed at a relatively large angle, although ID830 shows some blazar-like properties such as a one-sided jet and a core-dominated spectrum.

\subsection{Most distant radio quasar in super-Eddington phase with resolved core--jet}
Figure~\ref{fig:z_vs_lambdaEdd} summarizes the location of ID830 in the plane of Eddington ratio and $z$, as well as other radio sources with resolved parsec-scale core--jet structures observed with VLBI from the literature. The sample includes various classes of AGN, such as radio-loud quasars, radio galaxies, and NLS1s. Blazars are excluded from the sample because their optical/UV luminosities can be significantly affected by relativistic beaming and variability, making reliable and accurate estimates of the Eddington ratio difficult. As shown in Figure~\ref{fig:z_vs_lambdaEdd}, some nearby NLS1s are reported to exhibit jet activities associated with high- or super-Eddington accretion \citep[e.g.,][]{Doi2015,Hada2018,Yang2020,Wang2023a,Wang2023b}. In comparison with these sources, ID830 (red star) is the highest-redshift and currently the only known super-Eddington high-$z$ radio-loud quasar with an associated resolved jet. This suggests that super-Eddington accretion capable of launching jets, similar to radio-luminous NLS1s, can exist even in the early Universe. Our results also provide the first evidence that such a super-Eddington accretion phase can produce collimated jets extending over several hundred parsecs.

Although Figure~\ref{fig:z_vs_lambdaEdd} suggests that jetted NLS1s may be regarded as low-$z$ analogues of ID830 in terms of their high Eddington accretion rates, ID830 deviates significantly from this population in several physical properties. In NLS1s, high accretion rates close to or even exceeding the Eddington limit are generally associated with relatively low-mass black holes \citep[$\sim 10^5 \text{--} 10^{7.5}~\Msun$; e.g.,][]{Zhou2006,Yang2020}, whereas ID830 hosts a much more massive black hole of $\Mbh = 10^{8.64}~\Msun$, despite its extreme accretion rate of $\lambda_\mathrm{Edd,UV}=1.44$ and $\lambda_{\mathrm{Edd,X}} = 12.8$ \citep{Obuchi2026}. These high-$z$ and high-mass properties imply that the evolutionary track of ID830, indicating a history of remarkably rapid black hole growth, is substantially different from that of nearby NLS1s. 
ID830 also exhibits the 5~GHz radio luminosity of $L_\mathrm{5GHz} > 10^{43.3}$~erg~s$^{-1}$, which is much brighter than those of NLS1s \citep[$\lesssim 10^{41}$~erg~s$^{-1}$; e.g.,][]{Yang2020,Wang2023a,Wang2023b}.  
This suggests that super-Eddington accretion in ID830 can produce a more powerful radio jet, in contrast to the relatively weak jets typically observed in NLS1s.
Overall, ID830 may reveal a new class or evolutionary stage that is not well captured by existing samples of either sub-Eddington radio-loud quasars and super-Eddington NLS1s.

\section{Discussion}\label{sec:discussion}

Our observation indicates that ID830 hosts the relativistic, collimated jet extending up to 100-750~pc scale, and it is associated with the super-Eddington accretion phase.
The well-collimated jet of ID830 extends over $\approx 745$~pc at least at 1.6~GHz band, with the relativistic jet speed of $\gtrsim 0.19c$. Assuming a jet speed in the range of $0.19c$--$1c$, the jet launching timescale is estimated to be $10^{3.4}$--$10^{4.1}$~yr. This timescale is consistent with the typical episodic lifetime of quasar ($10^3$--$10^5$~yr; \citealt{Schawinski2015,Shen2021,Pflugradt2022}).
If this jet is associated with the current super-Eddington phase, the jet activity in ID830 does not originate from a short-lived burst driven by external factors lasting for only a few years (e.g., TDEs; \citealt{Bloom2011,Burrows2011,Wu2018}), but with a sustained super-Eddington accretion phase lasting for at least $10^{3-4}$~yr. Additionally, such a high radio luminosity of ID830 implies a jet kinetic power of $P_\mathrm{jet} > 10^{45.4}$~erg~s$^{-1}$, which may be sufficient for effective AGN feedback in the host galaxy \citep{McNamara2007}\footnote{The jet kinetic power is estimated using the relation between $P_\mathrm{jet}$ and $L_\mathrm{1.4GHz}$ \citep{Cavagnolo2010}: $P_\mathrm{jet} = 1.05 \times 10^{44}~(L_\mathrm{1.4\,GHz} / 10^{24}~\mathrm{W\,Hz^{-1}})^{0.75}\,\mathrm{erg\,s^{-1}}$. We note, however, that the $P_\mathrm{jet}$ and $L_\mathrm{1.4\,GHz}$ relation comes with several caveats, including the time evolution of $L_\mathrm{1.4\,GHz}$, large scatter in the slopes, and dependence on additional parameters. Therefore, the discussion above is based on an approximate estimator (see also \citealt[]{Igo2024}).}. We note that further direct evidence of jet--ISM coupling or star formation suppression is required to assess the impact of AGN feedback.

Magnetohydrodynamic simulations have shown that strong outflows can be launched in super-Eddington accretion flows on scales of a few gravitational radii ($r_\mathrm{g}$) \citep[e.g.,][]{Ohsuga2009,Ohsuga2011,Sadowski2015,Jiang2019}. These outflows are produced with velocities of $\sim 0.1c$--$0.4c$ and are subsequently collimated by optically thick gas up to scales of $\sim 100$--$200r_\mathrm{g}$ \citep{Jiang2019}. However, our observational results require a collimated, coherent jet or outflow extending to scales of hundreds of parsecs, or even kiloparsecs. This implies that an additional collimation or confinement mechanism, possibly related to magnetic fields as suggested for classical radio jets, must operate far beyond the launching region. The physical process that sustains such large-scale jet collimation in super-Eddington accretion remains a missing piece, calling for further theoretical and numerical studies. If such jet formation is shown to commonly occur in the super-Eddington regime, this would suggest that AGN feedback can be effective even during the rapid growth phase of SMBHs in the early Universe.


\acknowledgments
We appreciate the anonymous referee for a very careful reading of the manuscript and numerous helpful suggestions that greatly strengthened the paper.
This work is supported by the Japan Society for the Promotion of Science (JSPS) KAKENHI (26K21725 and 25H00660, K.~Hada; 25K01043, K.~Ichikawa).
K.H. also acknowledges support from Daiko Foundation (grant J0SE807004), the University Research Support Grant from the National Astronomical Observatory of Japan (NAOJ), and Grant-in-Aid for Outstanding Research Group Support Program in Nagoya City University Grant Number 2530002. 
K.I. also acknowledges support from the JST FOREST Program, Grant Number JPMJFR2466 and the Inamori Research Grants, which helped make this research possible. 
This research has made use of data from the MOJAVE database that is maintained by the MOJAVE team \citep{Lister2018}.

%

\vspace{3mm}
\facilities{VLBA, VLA, eROSITA}





\vspace{12mm}
\appendix

\section*{Robustness of the Jet Detection with $uv$-Tapered Images}\phantomsection\label{app:tapered_images}

\begin{figure*}[tp!]
\begin{center}
\includegraphics[width=0.75\textwidth]{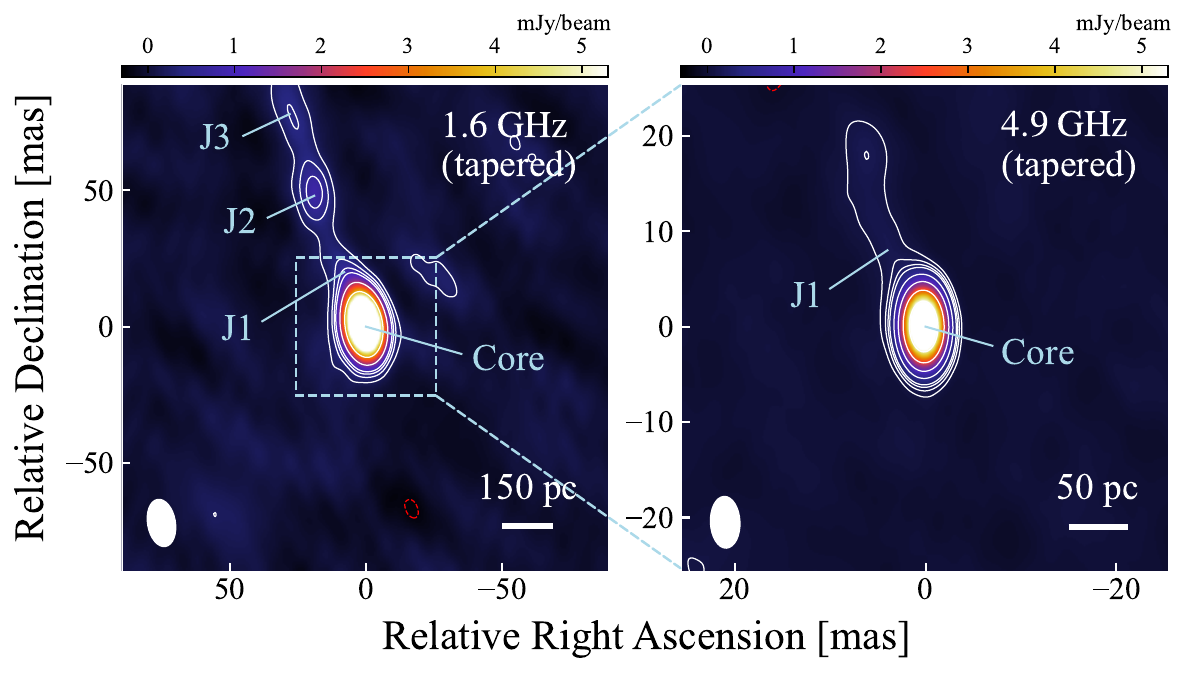}
\caption{
The $uv$-tapered VLBA images of ID830 at 1.6~GHz (left) and 4.9~GHz (right). The negative and positive contours are shown as the red dashed and solid white lines, respectively. The contour levels are at $-1$, 1, 2, 3, 4, 8, ..., $2^n$ times the $3 \sigma$ rms noise level. The white ellipse in each panel represents the beam size. The core and jet components (J1--J3) are also labeled in the figure. We adopt a different color scale from that used in Figure~\ref{fig:VLBAimages} to better highlight the jet emission.
}\label{fig:VLBA_tapered_images}
\end{center}
\end{figure*}

Our VLBA observations resolve a core--jet structure of ID830 at 1.6 and 4.9~GHz, as shown in Figure~\ref{fig:VLBAimages}. 
As an additional test of the robustness of the jet detection, we create $uv$-tapered images using the \verb|uvtaper| task in DIFMAP. The \verb|uvtaper| task applies a Gaussian taper to the visibility data, specified by a taper weight and a $uv$ radius (in M$\lambda$), to increase sensitivity to low-surface-brightness emission. We fix the taper weight to 0.3 and check the images at several $uv$ radii, confirming that the jet peak flux density and the jet S/N improve as the long-baseline visibilities are down-weighted. Figure~\ref{fig:VLBA_tapered_images} shows representative $uv$-tapered images at both 1.6 and 4.9~GHz, obtained using a taper corresponding to $\sim 1/3$ of the longest baseline. At 1.6~GHz (left panel), with the \verb|uvtaper| parameter of (0.3, 20), the jet S/N is improved by a factor of $\sim 1.2$ compared to the non-tapered image, reaching $>11 \sigma$. At 4.9~GHz (right panel), with the \verb|uvtaper| parameter of (0.3, 60), the jet S/N is improved by a factor of $\sim 1.1$ compared to the non-tapered image, reaching $>6 \sigma$. These results further support the robustness of our jet detection.

\bibliographystyle{aasjournal}
\bibliography{eFEDSradioVLBA}



\end{document}